\documentstyle[prl,aps,multicol]{revtex}
\input{epsf}
\begin{document}
\draft
\title{Quantum Poincar\'e Recurrences for Hydrogen Atom 
in a Microwave Field} 
\author{Giuliano Benenti$^{(a)}$, Giulio Casati$^{(b,c,d)}$, 
Giulio Maspero$^{(e)}$, and Dima L. Shepelyansky$^{(f)}$} 
\address{$^{(a)}$CEA, Service de Physique de l'Etat Condens\'e, 
Centre d'Etudes de Saclay, F-91191 Gif-sur-Yvette, France}  
\address{$^{(b)}$International Center for the Study of Dynamical 
Systems, Via Lucini 3, \\
Universit\`a degli Studi dell'Insubria, 22100 Como, Italy} 
\address{$^{(c)}$Istituto Nazionale di Fisica della Materia, 
Unit\`a di Milano, Milano, Italy} 
\address{$^{(d)}$Istituto Nazionale di Fisica Nucleare, 
Sezione di Milano, Milano, Italy} 
\address{$^{(e)}$Universit\`a Pontificia della S. Croce,   
Piazza di S. Apollinare 49, 00186 Roma, Italy} 
\address{$^{(f)}$Laboratoire de Physique Quantique, UMR 5626
du CNRS, Universit\'e Paul Sabatier, 31062 Toulouse, France} 
\date{\today}
\maketitle

\begin{abstract}
We study the time dependence of the ionization probability of 
Rydberg atoms driven by a microwave field, both in classical 
and in quantum mechanics. 
The quantum survival probability follows the classical one 
up to the Heisenberg time and then decays algebraically  
as $P(t)\propto 1/t$. 
This decay law derives from the exponentially long times required to   
escape from some region of the phase space, due to tunneling 
and localization effects.   We also provide
parameter values which should allow to observe such  decay in laboratory 
experiments. 
\end{abstract}
\pacs{PACS: 32.80.Rm, 05.45.Mt}  

\begin{multicols}{2}
\narrowtext

During the last two decades the manifestations of classical chaos in 
microwave ionization of Rydberg atoms have been studied experimentally 
by different groups and many interesting results have been obtained   
\cite{Koch,Bayfield,Walther,Kochrep}. In particular, laboratory 
experiments showed the quantum suppression of classically 
diffusive ionization process, in agreement with the predictions of 
dynamical localization theory \cite{IEEE}. 
The experimental technique based on accelerated proton beams, which 
is used for production of hydrogen atoms, 
allows to obtain interaction times with 
the microwave field of only few hundreds microwave periods 
\cite{Koch,Bayfield,Kochrep}. On the contrary, the thermal beams
used with alkali Rydberg atoms allow to vary the interaction time by 
orders of magnitude up to $10^5$ microwave periods \cite{Walther}. 
The first experiments of  Walther's group \cite{Walther} indicated 
an anomalously slow decay of the $10\%$-ionization threshold field  
as a function of the interaction time. This result cannot be explained  
within the picture of diffusive ionization in the domain of 
classical chaos. Some suggestions have been put forward to 
explain this slow decay which was attributed to some possible 
effects of noise for such long interaction times \cite{Smilansky,FS91}. 
More recently, new experimental data for the behavior of the survival  
probability $P(t)$ with time have been presented \cite{Delande}, showing an 
algebraic law decay $P(t)\propto t^{-\alpha}$, with $\alpha\approx 0.5$. 
In the same paper, numerical simulations of quantum dynamics have been 
made, giving a value of $\alpha$ consistent with experimental 
data. The origin of the slow algebraic decay was attributed to the 
underlying structure of classical mixed phase space composed by 
integrable islands surrounded by chaotic components. 
However, the investigations of classical chaotic systems with mixed 
phase space showed that the probability of Poincar\'e recurrences to 
the same region, or the survival probability up to time $t$, 
decays algebraically with power $\alpha\approx 1.5-3$ \cite{CS99,geisel,hclas}. 
Moreover, since the integral $\int_{t}^\infty P(\tau) d\tau$ is  
proportional to the measure of the finite chaotic region
where the trajectory is trapped, the value of $\alpha$ should be 
greater than one. According to the correspondence principle, one expects 
that, in the semiclassical regime, classical and quantum systems exhibit 
the same decay law.   
Therefore the above exponent $\alpha\approx 0.5$ 
found in the experiments requires, in this respect, an explanation. 
In particular the question arises whether this obtained value is 
generic or corresponds to some initial transient time behavior in a 
regime where quantum effects play an important role. 

Recent studies of quantum Poincar\'e recurrences for 
the Chirikov standard map 
in the semiclassical regime with mixed phase space \cite{QPR}
showed that quantum $P(t)$ follows 
the classical decay during a relatively large time $t_H$.
The time $t_H$ gives the Heisenberg time scale, which is determined by 
inverse level spacings. For $t>t_H$, the quantum survival probability 
starts to decay inversely proportional to time ($\alpha=1$) and becomes 
much larger than the classical one. The power $\alpha=1$ is due to 
exponentially long times required to escape from some region of phase 
space\cite{QPR}. These  exponentially long escape times are originated by tunneling 
from classically integrable region or by the exponential quantum localization. 
The above quantum behavior, with exponent $\alpha=1$, is different 
from the experimental data \cite{Delande} and this constitutes an 
additional motivation for the present paper. 
Indeed the highly excited states of hydrogen atom in a microwave field 
can be described by the Kepler map which is very similar to the Chirikov
standard map \cite{IEEE} and therefore one would expect the 
same behavior for the time dependence of survival probability. 

In order to investigate the probability decay for the hydrogen atom in a 
microwave field, we choose the initial state with principal quantum number 
$n_0$ and numerically studied the survival probability $P(t)$ in a 
linearly polarized monochromatic electric field $\epsilon(t)=
\epsilon\sin(\omega t)$. Here $\epsilon$ and $\omega$ are the 
strength and frequency of the microwave field, measured in atomic 
units. 
The quantum evolution is numerically simulated by the Kepler 
map \cite{IEEE}, by the one-dimensional ($1d$) model of a hydrogen atom and by the 
$3d$ model for atoms initially prepared in states extended along  
the field direction and with magnetic quantum number $m=0$. 
The comparison of these three models  
shows that the essential physics is captured by the $1d$ model as 
already discussed in \cite{IEEE}. In addition we show that also the
Kepler map gives an approximate correct description of the dynamics. 

The hydrogen atom in a linearly polarized monochromatic electric field 
is described by the Hamiltonian 
\begin{equation} 
H=\frac{p^2}{2}-\frac{1}{r}+\epsilon z \sin (\omega t),
\end{equation} 
where, in the $1d$ model the motion is assumed to take place along the 
field direction ($z$-axis, with $z\geq 0$). 
In order to compare classical and quantum dynamics it is convenient 
to use the scaled field strength $\epsilon_0=\epsilon n_0^4$ and frequency 
$\omega_0=\omega n_0^3$, which completely determine the classical 
dynamics. The classical limit corresponds to $\hbar_{\rm eff}=   
\hbar/n_0\to 0$, at constant $\epsilon_0$, $\omega_0$. 
For $\omega_0>1$ the main change of the electron 
energy $E$ occurs when the electron is close to the nucleus. 
As a consequence the dynamics is approximately given by 
the Kepler map\cite{IEEE}. 
\begin{equation} 
\overline{N}=N+k\sin\phi, \quad 
\overline{\phi}=\phi+2\pi\omega(-2\omega\overline{N})^{-3/2},
\label{kmap} 
\end{equation} 
where $N=E/\omega$, $k=2.6\epsilon 
\omega^{-5/3}$, $\phi=\omega t$ is the phase of the microwave 
field when the electron passes through the perihelion
and the bar marks the new values of variables.
In the quantum case, the change of $N$ gives the number of absorbed 
photons while the number of photons required to ionize the atom is  
$N_I=1/(2n_0^2\omega)$. In classical mechanics diffusive ionization  
takes place for fields above the chaos border: 
$\epsilon_0>\epsilon_c\approx 1/(49\omega_0^{1/3})$ 
\cite{IEEE}. 
The quantum dynamics of the model (\ref{kmap}) is described by the 
quantum Kepler map for the wave function $\psi(\phi)$: 
\begin{equation} 
\overline{\psi}=\exp(-iH_0(\hat{N}))
\hat{P}\exp(-ik\cos\hat{\phi})\psi,
\end{equation} 
where $H_0(\hat{N})=2\pi/\sqrt{-2\omega\hat{N}}$, 
$\hat{N}=-id/d\phi$, $\hat{\phi}=\phi$ ($-\infty<\phi<+\infty$),  
and the operator $\hat{P}$ projects probability over the states 
with negative energy ($N<0$) \cite{IEEE}. 
We introduce an absorption border 
for levels with $n\geq n_c$, which for the Kepler map corresponds to 
$N\geq N_c\approx -1/(2n_c^2\omega)$ \cite{note}. 
Such  border occurs in real laboratory experiments, for example as 
a consequence of unavoidable static electric  field experienced 
by the Rydberg atoms during their interaction with the microwave field.
The absorption border $n_c$ can be varied in a controlled way via a 
static electric field $\epsilon_s$, the static field ionization 
border being $\epsilon_s n_c^4\approx 0.13$. 

\begin{figure}
\centerline{\epsfxsize=8cm\epsffile{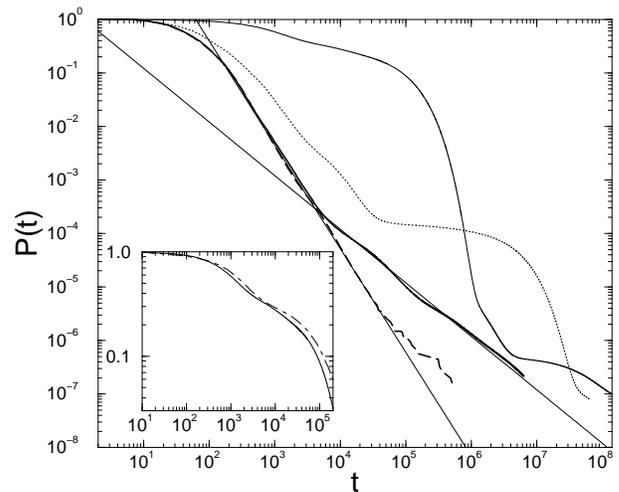}}
\caption{Survival probability $P(t)$ as a function of the interaction 
time $t$ (in units of microwave periods) for $\epsilon_0=0.065$, 
$\omega_0=1.6$, $n_c=2 n_0$: quantum Kepler map for $n_0=23$ 
(thin solid line), $n_0=230$ (thin dotted line), $n_0=2300$ 
(thick solid line) and classical Kepler map (thick dashed line, 
ensemble of $10^9$ trajectories). The straight lines have slopes 
$1$ and $1.94$, the latter coming from a fit of the classical 
decay for $2\times  10^2<t<3\times 10^4$. Inset: quantum Kepler 
map (solid line) versus $1d$ hydrogen model (dot-dashed line) 
for $n_0=23$.}
\label{fig1} 
\end{figure}

The results of quantum simulations for 
the situation similar to the experimental one
 (Fig. 2 (b) in Ref. 
\cite{Delande}) are shown in Fig. \ref{fig1}. 
The Kepler map description allows us to study the quantum dynamics up 
to very long times ($t=10^8$, here and below 
time is given in microwave periods). 
In the case $n_0=23$ the quantum data for the survival probability 
$P(t)$ obtained from the quantum Kepler map and the $1d$ hydrogen 
atom model agree with each other (see the inset of Fig. \ref{fig1}) 
and with the numerical computations 
of \cite{Delande}. However all these data are strongly different 
from the classical probability decay shown in Fig.\ref{fig1}, which  
displays a slope $\alpha\approx 2$. 
The reason of this disagreement should be attributed 
to the fact that $n_0=23$ is not in the semiclassical regime.  
Our data for the Husimi distribution, obtained from the Wigner 
function by smoothing over the size $\hbar$ \cite{Jensen},  
show that a significant part of the probability is trapped 
inside the stable island at $n\approx 20$ ($\omega n^3\approx 1$). For this  
reason the probability decays slowly during a long time $t\approx 10^5$ 
after which it drops faster. 
If $n_0$ is increased significantly, the semiclassical regime is reached 
and the quantum probability decay becomes close to the classical one 
up to the time scale $t_H\approx 10^4$. Our data  
show that $t_H$ is proportional to $n_0$ (at fixed $\epsilon_0$, 
$\omega_0$), in agreement with previous estimates of Ref. \cite{QPR}, 
according to which $t_H\propto 1/\hbar_{\rm eff}$. After this time the 
quantum $1/t$ decay is clearly observed in agreement with the results 
of \cite{QPR}. 

\begin{figure} 
\centerline{\epsfxsize=8cm\epsffile{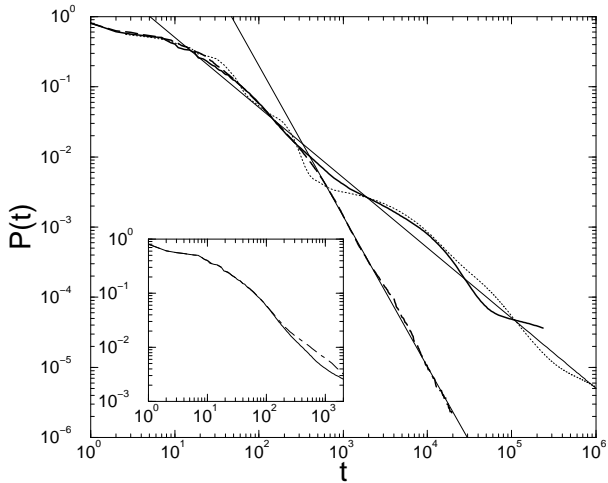}}
\caption{Survival probability for $\epsilon_0=0.1$, 
$\omega_0=2.6$, $n_0=60$, $n_c=64$: quantum Kepler map 
(dotted line), quantum (solid line) and classical (dashed 
line, ensemble of $3\times 10^6$ trajectories) 
$1d$ hydrogen model. The straight lines have slopes 
$1$ and $2.15$, the latter coming from a fit of the classical 
decay for $5\times  10^2<t<2\times 10^4$. Inset: quantum  
survival probability for the $1d$ model (solid line) and the 
$3d$ model (dot-dashed line).}  
\label{fig2} 
\end{figure}

In Fig. \ref{fig2} we show a more realistic case in which, initially, 
classical and quantum probabilities decay in a very similar way and 
where only after a time $t_H\approx 5\times 10^2$, the quantum survival probability 
starts to decay more slowly ($P(t) \propto 1/t$) than the classical one 
which decays as $1/t^{\alpha}$, with $\alpha\approx 2.15$. 
This case corresponds 
to $n_0=60$ and can be observed in experiments similar to those performed 
in \cite{Delande}. Again the quantum Kepler map gives a qualitatively 
correct description of the ionization process up to very long 
interaction times. The comparison of quantum simulations for the $1d$ hydrogen 
atom model and  the $3d$ dynamics is shown in the inset of Fig. \ref{fig2}. 
It demonstrates that both dynamics give very close results, confirming  
that the essential physics is captured by the $1d$ model.   
We put the absorption border  near  the initial state ($n_c=64$) in 
order to have $\rho_c=\ell_\phi/\Delta N_c \approx 3.5 > 1$, where
$\ell_\phi=3.3\epsilon_0^2 \omega_0^{-10/3} n_0^2$  is the localization length 
in number of photons \cite{IEEE} and 
$\Delta N_c=(n_0/2 \omega_0)(n_0^2/n_c^2-1)$  is the number of photons 
required to reach the absorption border. 
In this  way the probability can 
go out very easily and the $1/t$ probability decay is observed after a 
short transient time of the order of $20$ microwave periods. 
On the contrary, when $\rho_c <1$, as in the case of Fig. \ref{fig1} 
for $n_0=23$ ($\rho_c\approx 0.3$), strong fluctuactions around the 
$1/t$ decay take place. This is analogous to the huge 
(log-normally distributed) conductance fluctuactions 
in a disorder solid with localization length smaller 
than the system size \cite{pichard}.  

\begin{figure} 
\vspace{-0.5cm}
\centerline{\epsfxsize=4.2cm\epsffile{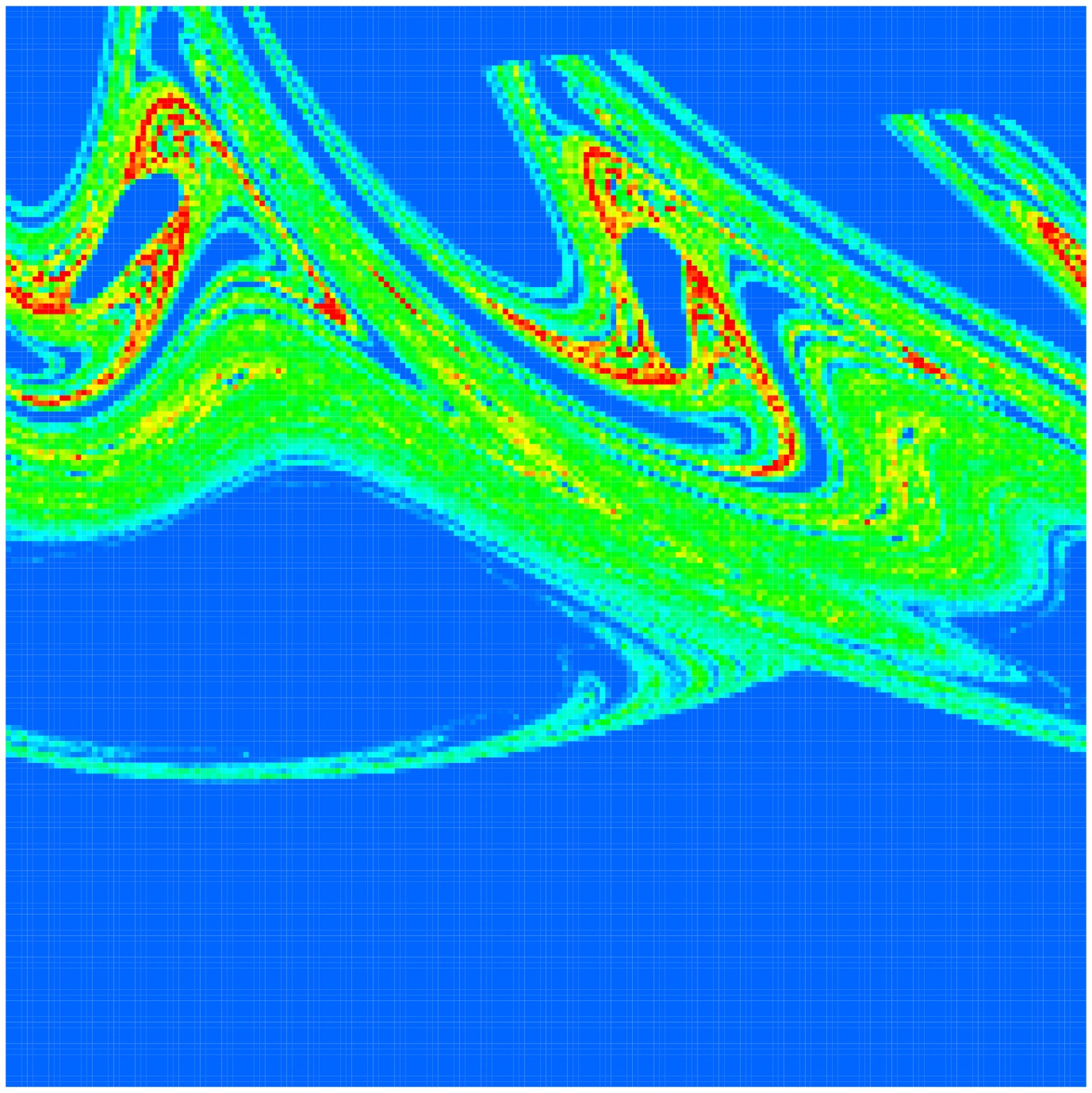}
\hfill\epsfxsize=4.2cm\epsffile{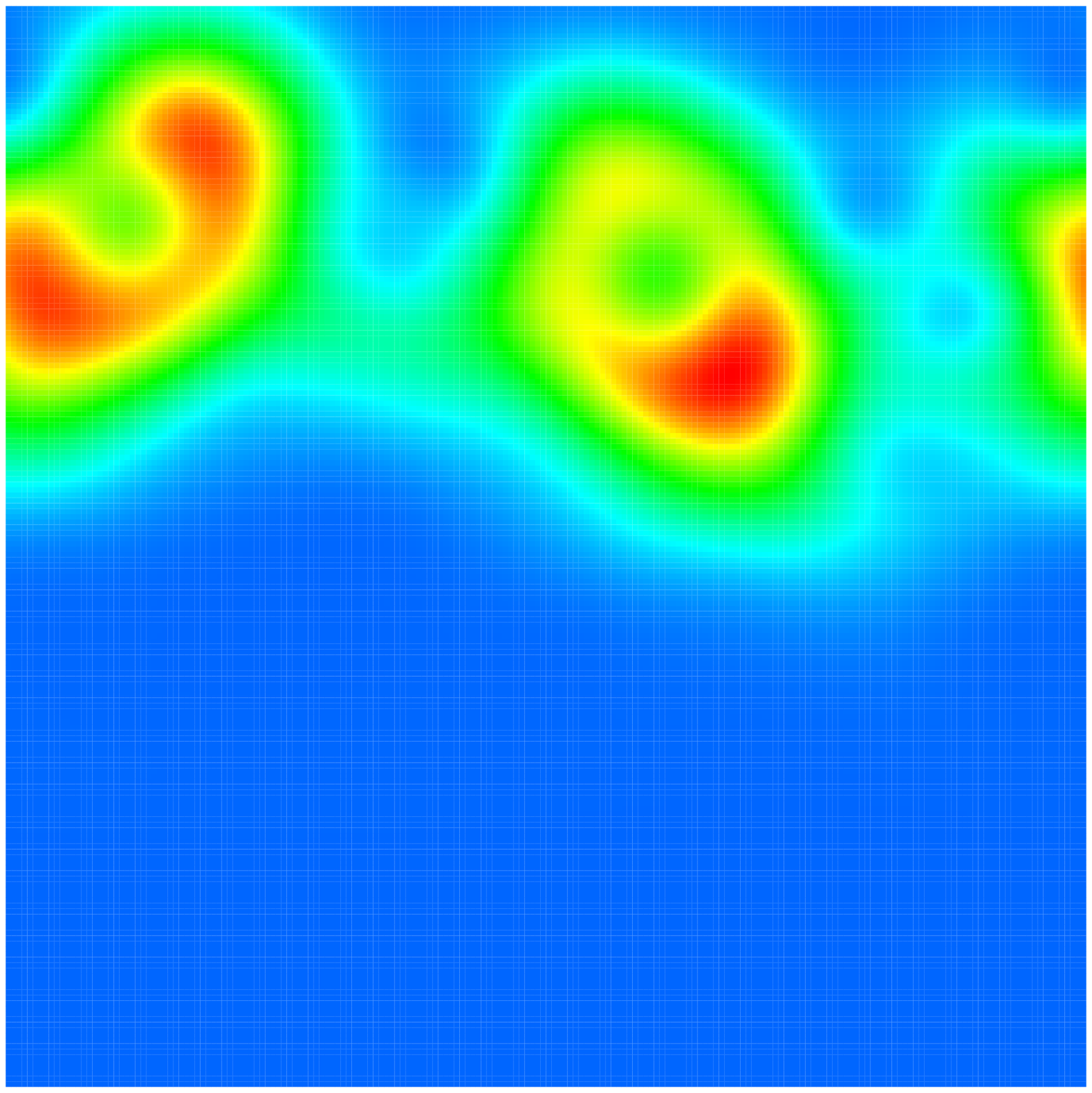}}  
\vspace{-1.55cm}
\centerline{\epsfxsize=4.2cm\epsffile{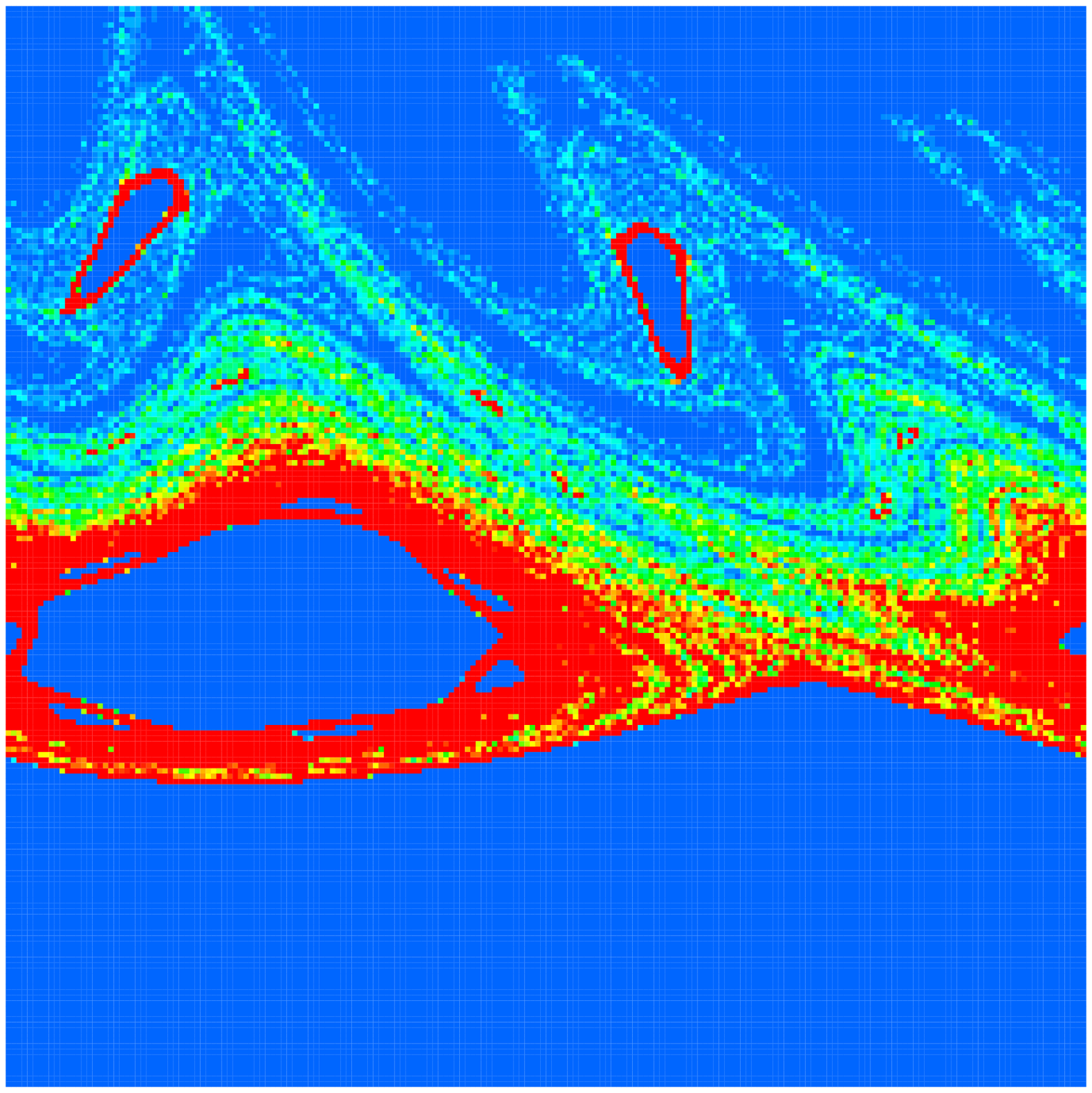}  
\hfill\epsfxsize=4.2cm\epsffile{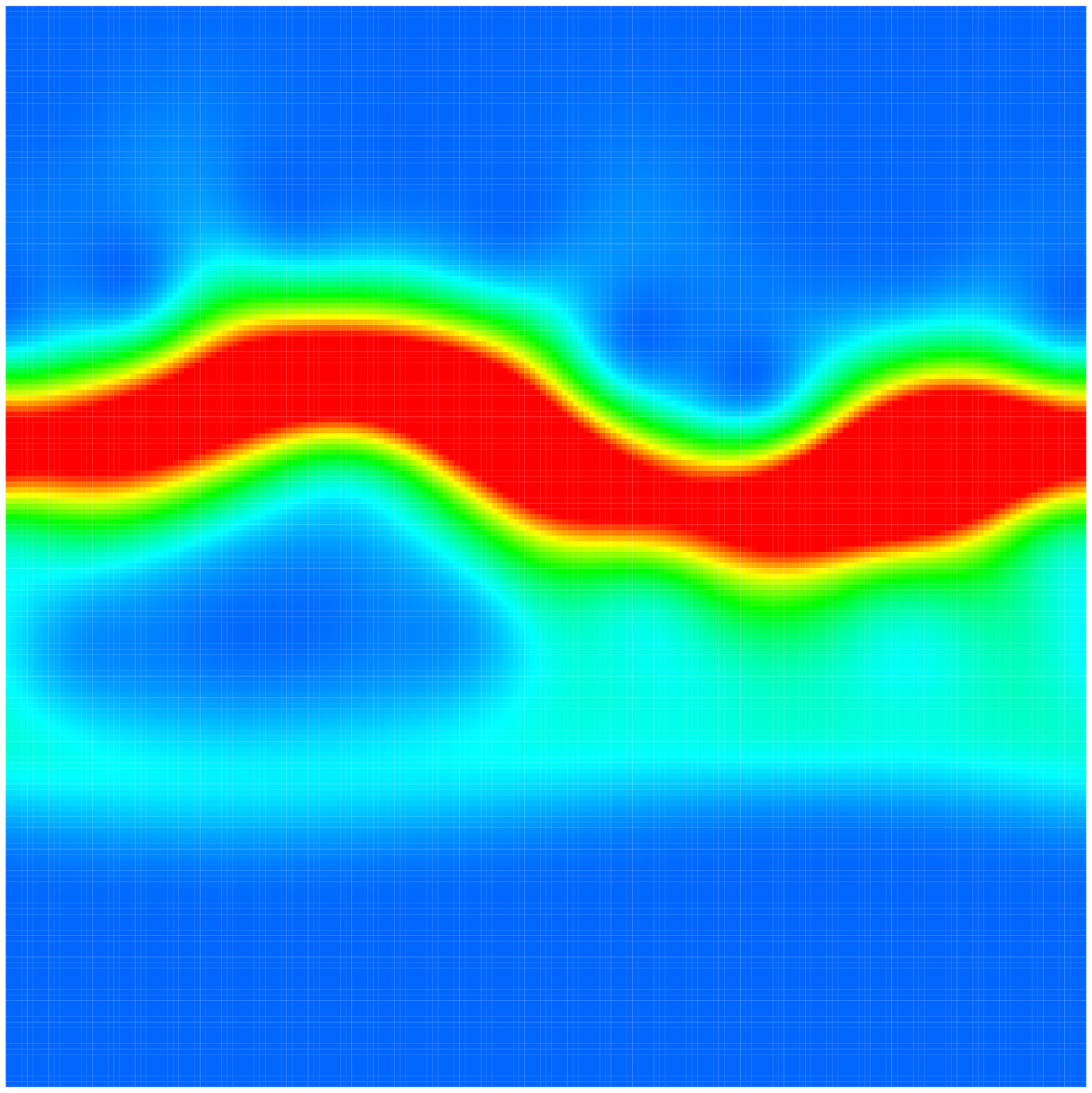}}   
\caption{Classical density plot (left) and Husimi function 
(right) in action-angle variables ($n,\theta$), with 
$30\leq n\leq 63.5$ (vertical axis) and 
$0\leq\theta<2\pi$ (horizontal axis),   
for the $1d$ model in the case of Fig. 2.  
Husimi function is averaged in a finite time
interval to decrease fluctuations: $50\leq t \leq 60$ (top); 
$2\times 10^3\leq t \leq 10^4$ (bottom left); 
$9.9\times 10^3\leq t \leq 10^4$ (bottom right).
The color is proportional to the density: blue for zero and 
red for maximal density.}  
\label{fig3} 
\end{figure}

In order to confirm that the algebraic probability decay is related 
to the sticking of classical trajectories and of quantum 
probability near the integrable islands in the phase space, we show 
in Fig. \ref{fig3} the time evolution of the survival probability 
distribution in the phase space of action-angle variables $(n,\theta)$
for the $1d$ model. 
In the classical case $3\times 10^6$ orbits were initially homogenously 
distributed in the angle $\theta$ on the line $n=n_0=60$, corresponding 
to the initial quantum state with principal quantum number $n_0=60$. 
After $50$ microwave periods, the classical distribution of 
non ionized orbits shows a fractal structure which surrounds 
the stability islands (Fig. \ref{fig3} top left). 
At larger times this distribution approaches 
more and more closely the boundary critical invariant curves
(Fig. \ref{fig3} bottom left). 
One of them confines the motion in the region with 
$n>n_b\approx n_0 (\epsilon_c/\epsilon_0)^{1/5} \approx 41$
where $n_b$ determines the classical chaos border
for given $\epsilon_0$. 
Other invariant curves mark the critical boundaries around internal 
stability islands (for example at $n\approx 55$, corresponding to 
$\omega n^3\approx 2$).   
In the quantum case the value of $\hbar_{\rm eff}$ is not sufficiently 
small to resolve the fractal structure at small scales. 
However, the Husimi function shows similarities with classical 
probability distribution at $t=50$ (Fig. \ref{fig3} top right). 
At longer times, the diffusion towards the boundary at $n_b$ is 
slowed down due to localization effects and penetration of the 
quantum probability inside the classical integrable islands.  
At $t=10^4$ (Fig. \ref{fig3} bottom right) the quantum probability 
is concentrated in a layer near $n_b$. Due to localization effects, 
the Husimi function does not change significantly for a very long 
interaction time ($10^3< t < 3\times 10^4$). Eventually  the
probability starts to penetrate very slowly inside the main island 
at $n\approx n_b$. 
Therefore tunneling and localization effects are 
responsible for the slow $1/t$ decay of the quantum survival probability 
seen in Fig. \ref{fig2}. 

\begin{figure} 
\vspace{-0.5cm}
\centerline{\epsfxsize=4.2cm\epsffile{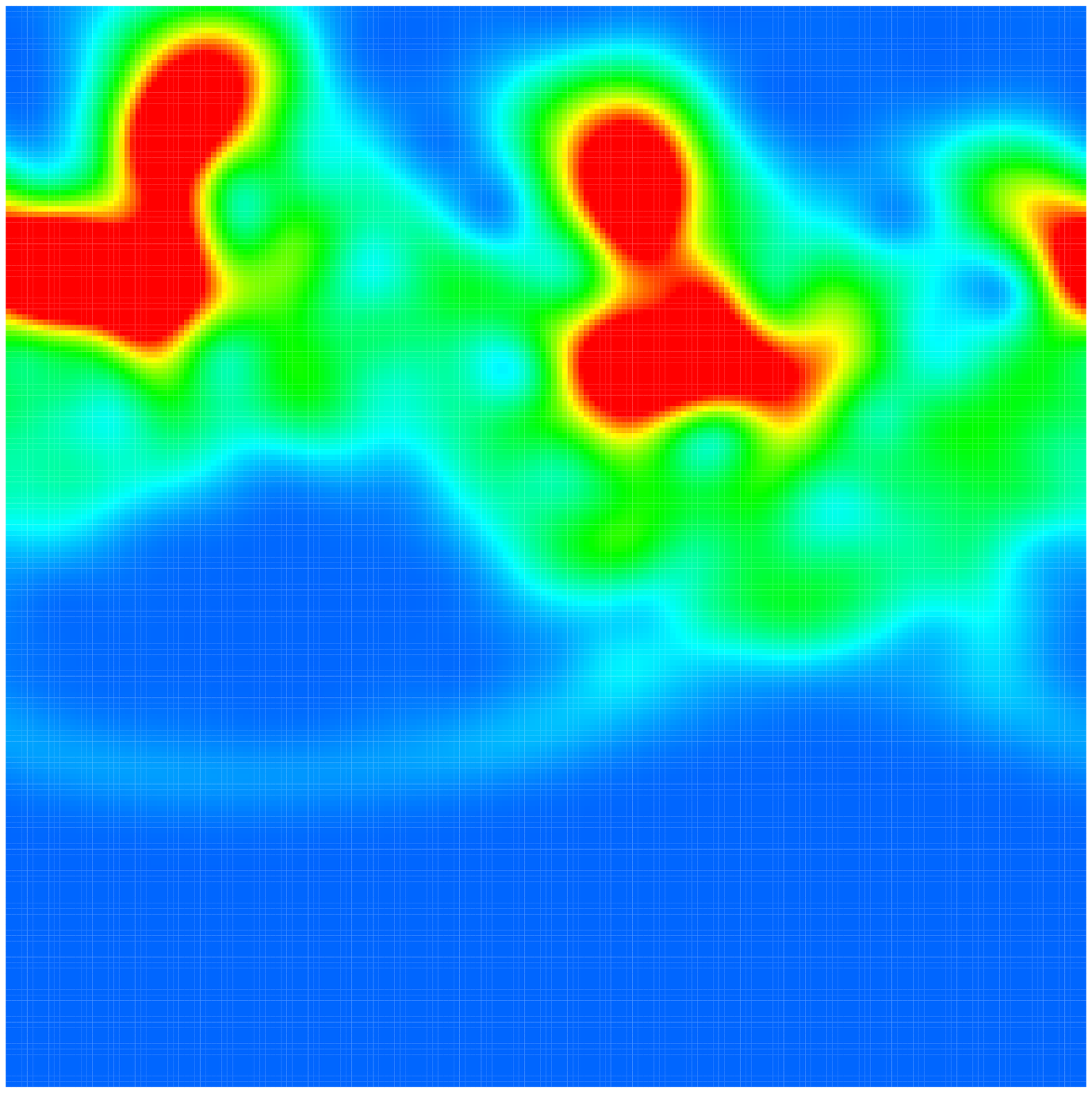}
\hfill\epsfxsize=4.2cm\epsffile{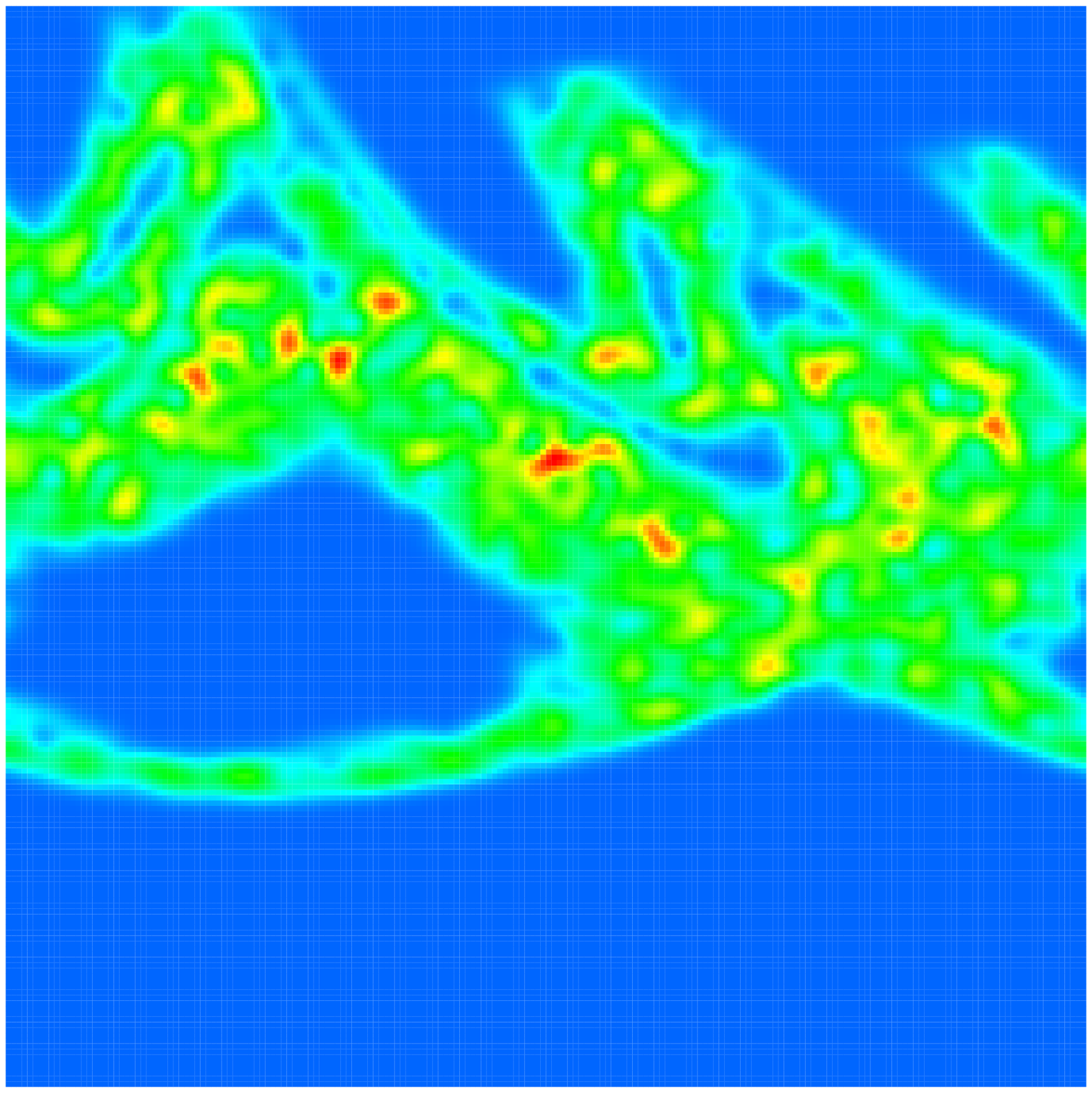}}  
\caption{Quantum fractal Husimi function for parameter values and interaction 
times as in Fig. \ref{fig3} top left, 
with $n_0=150$ (left) and $n_0=1200$ (right).} 
\label{fig4} 
\end{figure}

The fractal structure of the classical distribution is washed out 
at scales smaller than the minimal quantum cell $\hbar_{\rm eff}$. 
Therefore a better resolution can be obtained increasing the principal 
quantum number $n_0$, at fixed $\epsilon_0$, $\omega_0$, and $n_c/n_0$. 
The Husimi function clearly reflects the underline fractal structure 
at very high principal quantum numbers $n_0=150$ (Fig. \ref{fig4} left) 
and $n_0=1200$ (Fig. \ref{fig4} right). Similar {\it quantum fractals} 
have been found in the kicked rotator model with absorbing boundary 
conditions \cite{fractal}. 

Notice that the probability decay $P(t)$ is related to correlations 
decay via $C(t) \propto t P(t) $ \cite{CS99}. 
In the case of $\alpha=1$ this implies that correlations do not 
decay. The Fourier transform of $C(t)$ gives 
the spectral density $S(\omega)$ of the effective noise produced by the 
dynamics: $S(\omega)=\int C(t)\exp(i\omega t)dt\approx 
1/\omega$. This shows that the spectral noise associated with the quantum 
Poincar\'e recurrences with $\alpha=1$ scales like 
$S(\omega) \propto 1/\omega$. 
A similar behavior of noise has been observed in many 
scientific disciplines \cite{Press}, for example in the  
resistence fluctuactions of different solid state devices \cite{noise}. 
This phenomenon is known as $1/f$ noise and usually extends over  
several orders of magnitude in frequency, indicating a broad distribution of 
time scales in the system. In the case of quantum Poincar\'e 
recurrences this property stems from the exponentially low escape 
rate from some regions of the phase space. 

In summary, on the basis of our previous investigations 
and of the numerical studies presented in this paper we 
conclude that the survival probability for Rydberg atoms in a  
microwave field decays, up to the time scale $t_H\propto n_0$, in a way 
similar to the classical probability.  
For $t>t_H$ the quantum probability starts to 
decay slower than the classical one, with the exponent of the 
algebraic decay $\alpha=1$. 
We have given parameter values which should allow to observe  
quantum Poincar\'e recurrences  
in microwave experiments with Rydberg atoms. 

This research is done in the frame of EC program RTN1-1999-00400.

\end{multicols}


\begin{references}
\bibitem{note} We introduce complete absorption for levels 
with $n\geq n_c$ after each microwave period ($\Delta t=1$).
We checked that our results do not change significantly 
for a wide range of variation of $\Delta t$ around this value.
\end{references}
\end{document}